\documentclass{article}
\usepackage{spconf,amsmath,graphicx}

\usepackage{comment}
\usepackage{bm}
\usepackage{xcolor}
\usepackage{multirow}
\usepackage{makecell}
\usepackage{textcomp}
\usepackage{graphicx}

\usepackage{makecell}
\usepackage{longtable}
\usepackage{subcaption}
\usepackage{lipsum} 
\usepackage{multicol}

\usepackage{amsmath}
\usepackage{amssymb}

\title{AdaMER-CTC: Connectionist Temporal Classification with Adaptive Maximum Entropy Regularization for Automatic Speech Recognition}
\name{%
  \begin{tabular}[t]{@{}c@{}}
  SooHwan Eom$^{1}$, \quad Eunseop Yoon$^{1}$, \quad Hee Suk Yoon$^{1}$,\\
  Chanwoo Kim$^{2}$, \quad Mark Hasegawa-Johnson$^{3}$, \quad Chang D. Yoo$^{1}$\protect\textsuperscript{\dag} \thanks{\protect\textsuperscript{\dag}Corresponding author} \thanks{This work was partly supported by Institute of Information \& communications Technology Planning \& Evaluation (IITP) grant funded by the Korea government(MSIT) [NO.2022-0-00184, Development and Study of AI Technologies to Inexpensively Conform to Evolving Policy on Ethics] and SAMSUNG Research, Samsung Electronics Co., Ltd.}\\[1ex]
  \end{tabular}
}

\address{\textsuperscript{1} Korea Advanced Institute of Science and Technology, Daejeon, Republic of Korea \\ \textsuperscript{2} Korea University, Seoul, Republic of Korea \qquad \textsuperscript{3} University of Illinois Urbana-Champaign}

\begin{document}

\maketitle

\begin{abstract}
In Automatic Speech Recognition (ASR) systems, a recurring obstacle is the generation of narrowly focused output distributions. This phenomenon emerges as a side effect of Connectionist Temporal Classification (CTC), a robust sequence learning tool that utilizes dynamic programming for sequence mapping. While earlier efforts have tried to combine the CTC loss with an entropy maximization regularization term to mitigate this issue, they employed a constant weighting term on the regularization during the training, which we find may not be optimal. In this work, we introduce Adaptive Maximum Entropy Regularization (AdaMER), a technique that can modulate the impact of entropy regularization throughout the training process. This approach not only refines ASR model training but ensures that as training proceeds, predictions display the desired model confidence.
\end{abstract}
\section{Introduction}
\label{sec:intro}
Deep learning-based Automatic Speech Recognition (ASR) has significantly advanced due to the use of the Connectionist Temporal Classification (CTC) loss \cite{CTC}. CTC loss allows training on large datasets without requiring explicit alignment between speech-transcript pairs. Central to this alignment flexibility is the addition of the \textit{blank symbol} to the existing output label set, providing a mechanism to handle varying sequence lengths and consecutive identical characters. 

\begin{figure}[ht]
	\centering
	\includegraphics[width=1.0\linewidth,]{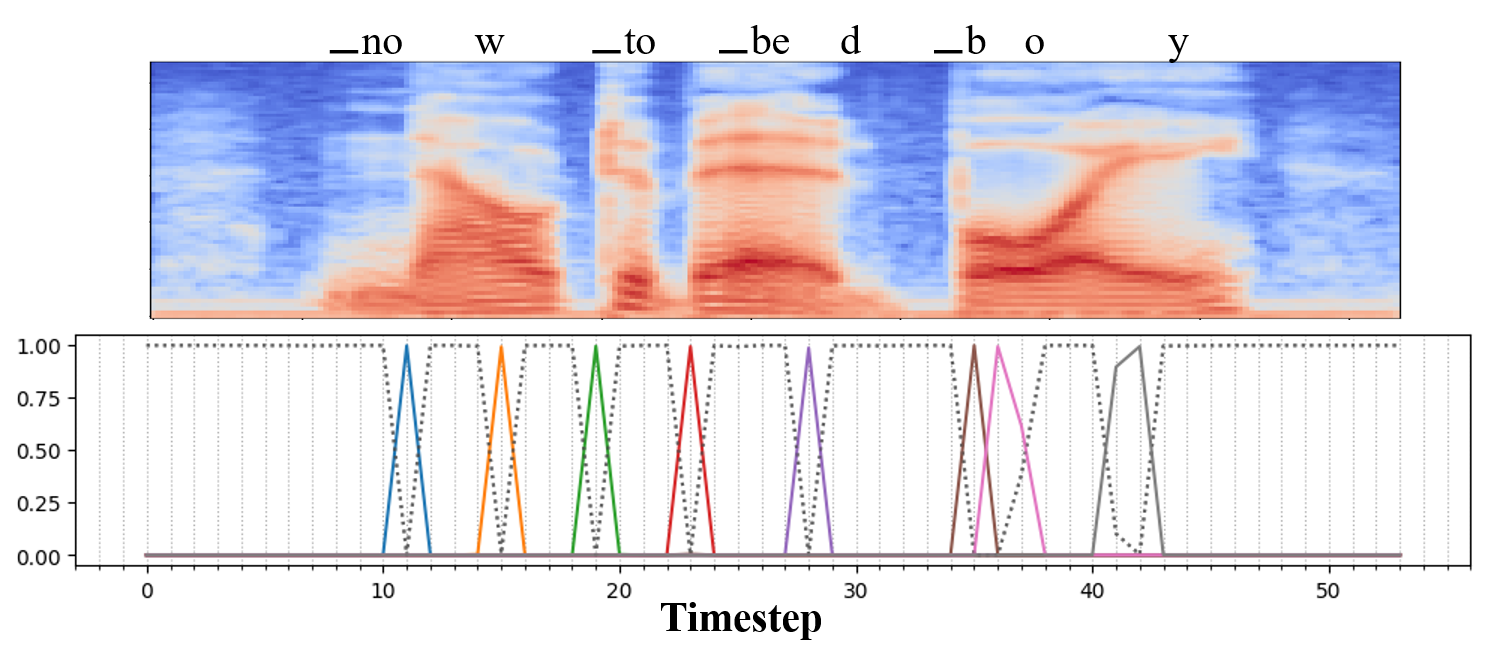}
	\caption{\textbf{Illustration of CTC peaky distribution output.} The dotted line denotes the output probability of the blank symbol.} 
	\label{fig:peaky}
\end{figure}

However, training with CTC loss often results in a peaky distribution as shown in Fig. \ref{fig:peaky}.
This peaky behavior is a manifestation of the blank symbol design and the entropy minimization property inherent to CTC loss \cite{peaky, peakblank1, peakblank2}. Randomly initialized ASR model exhibits a tendency to predict only blank symbols, as it searches for a single point to ``jump" to non-blank symbols in the target sequence, resulting in sharp spikes in its prediction. The entropy-minimizing nature of this training then leads to the predictions during training being stuck or converging into such peaky sub-optimal solutions, hindering the training process.


Prior works \cite{kim2018improved, ctc_entropy} have attempted to address this issue by combining the CTC loss with constant entropy maximization as an auxiliary regularization loss. However, our research suggests that such constant weighting on uncertainty might be counterproductive in the latter stages of training, as confidence in predictions becomes more crucial. In this paper, we propose Adaptive Maximum Entropy Regularization (AdaMER), which dynamically adjusts the effect of the entropy regularization throughout training. Through extensive experiments using the LibriSpeech \cite{librispeech} corpus, we show that AdaMER can effectively improve the CTC training of ASR models. 

\section{Related Work}
\label{sec:related_work}

\subsection{Overconfident Problem in CTC}
\label{sec:rel1}
The Connectionist Temporal Classification (CTC) framework \cite{CTC} has significantly impacted numerous end-to-end sequence learning tasks, revolutionizing areas such as speech recognition \cite{speech_recognition_1, speech_recognition_2}, text recognition \cite{text_recognition_1}, and video segmentation \cite{video_segmentation}, due to its ability to model the sequential output without the need for explicit alignment information between input and output sequences.
However, despite the substantial breakthroughs made possible by the use of CTC, there exist a few persisting challenges.

One such challenge is the tendency of the CTC-trained model to generate highly peaky distributions \cite{CTC, peaky, peakblank1}, a characteristic often interpreted as a sign of overconfidence.
This `overconfidence' refers to scenarios in which the model predicts certain outputs with high certainty, which, while positive if the prediction is correct, can lead to pronounced errors if the prediction is off the mark.
The `peaky behavior' of CTC is highly related to the 'blank' symbol, which is an additional label for handling consecutive identical labels. According to the analysis from \cite{peaky, peakblank2}, while blank symbols are crucial for CTC loss to ensure convergence, it is also the cause of the localized peaky predictions.
They specifically point out that models that are initialized uniformly and then trained using gradient descent methods—a common practice in machine learning—are susceptible to converging to suboptimal local optima. These suboptimal solutions are often marked by overconfidence or peaky behavior, thereby suggesting that the issue is intrinsically linked to the model's training dynamics rather than the model architecture itself.



\subsection{Regularization in CTC}


Model overconfidence is a common challenge across the machine learning landscape, prompting the creation of a multitude of regularization techniques to handle it. 
One solution to the overconfidence problem is to directly penalize the model confidence, which is often expressed as the entropy of the model prediction. Label smoothing regularization\cite{label_smoothing}, which uses soft targets instead of hard one-hot targets for cross-entropy labels, can be seen as maximizing relative entropy between the model prediction and uniform distribution\cite{entropy_related1, generalized_entropy}. In ASR, prior works including \cite{kim2018improved} have applied label smoothing on the CTC criterion.

Confidence regularization is also connected to the maximum entropy principle \cite{maxent_theory}, which states that the distribution that best describes the current state is the one that leaves the largest amount of uncertainty (or entropy) consistent with the constraint. Strategies such as maximum entropy regularization are widely used within the reinforcement learning domain \cite{max_ent_reinforcement1, SAC, SAC2}, which have been deployed to foster exploration and deter early convergence. 
For CTC, EnCTC \cite{ctc_entropy} have proposed a maximum conditional entropy-based regularization for CTC, applied to the optical character recognition task. This method considers the conditional distribution of potential paths given the input sequence and label sequence. Based on the fact that the error signal of CTC is proportional to the likelihood of the path, EnCTC is designed to prevent the entropy of feasible paths from declining too rapidly, thus mitigating the impact of CTC’s convergence to a single path and encouraging exploration during training. 

\begin{figure}[t]
	\centering
	\includegraphics[width=1.0\linewidth]{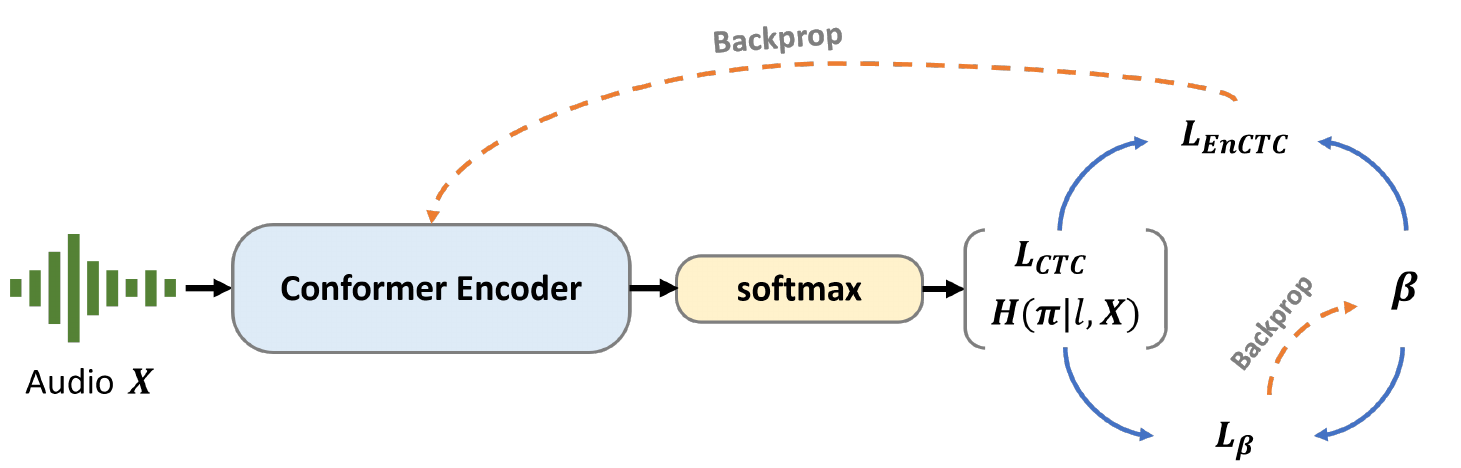}
	\caption{\textbf{CTC with Maximum Entropy Regularization.} An entropy-based regularization term is added to the original CTC objective to increase the entropy while training. This boosts the probability of the paths nearby, enhancing exploration throughout the training process.} 
	\label{fig:method}
\end{figure}

\section{CTC with Maximum Entropy Regularization}
\label{sec:method}

In this section we explain the standard CTC method, the maximum entropy regularization, and our AdaMER method.



\subsection{Problem Formulation} 
Consider a speech dataset $\mathcal{D}$. This dataset consists of pairs of input sequences and their corresponding target label sequences. A pair is denoted as $(X, l)$, where $X = [X_{1}, ..., X_{T}]$ is an input sequence of length $T$ and $l = [l_{1}, ..., l_{L}]$ is the corresponding target label sequence of length $L$, such that $L \leq T$. All possible elements of $l$ is inside the fixed-sized alphabet set $\mathcal{S}$. In this study, $X$ represents a speech sample, and $l$ is the corresponding transcription. The alignment between $X$ and $l$ is not known but in a sequential manner, which is a common scenario in many speech recognition problems.

\subsection{Connectionist Temporal Classification} 
Connectionist Temporal Classification (CTC) provides a way to handle the alignment problem in sequence prediction tasks \cite{CTC}. It introduces a new objective function, referred to as the CTC loss. The CTC loss is essentially the negative log-likelihood of the correct label sequence given the input sequence. This is mathematically represented as:
\begin{equation}
    \mathcal{L}_\text{CTC} = - \log p(l|X_{1:T})
\end{equation}



CTC defines the conditional likelihood $p(l|X_{1:T})$ as the sum of probabilities of all possible valid alignments between the input and label sequence.
Such alignment is called the path $\pi$, and the many-to-one mapping from the feasible paths to the label sequence $l$ is denoted as $\mathcal{B}(\pi)$. The conditional likelihood then can be written as:
\begin{equation}
    p(l|X_{1:T}) = \sum_{\pi \in \mathcal{B}^{-1}(l)}p(\pi | X_{1:T})
\end{equation}
and
\begin{equation}
    p(\pi | X_{1:T}) = \prod_{t=1}^{T} y_{\pi_t}^t
\end{equation}
where $y_k^t$ is the probability for symbol $k$ on timestamp $t$. This probability is usually calculated by the softmax function:
\begin{equation}
    y_k^t = \frac{e^{a_k^t}}{\sum_{k'}e^{a_{k'}^t}}
\end{equation}
where $\{a_k^t | t \in [1, T], k \in \mathcal{S}'\}$ is the output right before the softmax activation layer and $\mathcal{S}'=\mathcal{S} \cup {\emptyset}$ is the alphabet set including the additional `blank' symbol denoted as $\emptyset$. This additional `blank' label is introduced to handle varying lengths between the input sequence and the output sequence and also deals with the unknown alignment problem. The total number of output labels for CTC becomes $|\mathcal{S}|+1$.

In practical implementations, CTC uses a dynamic programming algorithm, akin to the forward-backward algorithm in Hidden Markov Models (HMM), to efficiently compute these probabilities. The final objective is to maximize the total probability of the correct label sequence over all possible alignments.

\subsection{Maximum Entropy Regularization} 

To resolve the overconfident peaky prediction, an entropy-based regularization method, named Entropy-Regularized Connectionist Temporal Classification (EnCTC), has previously been used for the optical character recognition task \cite{ctc_entropy}.
Maximum Entropy Regularization is a widely used method in reinforcement learning, such as in Soft Actor-Critic \cite{SAC, SAC2}, which encourages the agent to explore the broader policy space and prevents the agent from converging to sub-optimal policy. Inspired by this, EnCTC introduces a maximum conditional entropy-based regularization term which can be represented as follows:
\begin{equation}\label{eq.enctc}
    L_\text{EnCTC} = L_\text{CTC} - \beta H(\pi|l, X)
\end{equation}
where $\beta$ is a factor that regulates the intensity of the maximum conditional entropy regularization.

The entropy of the feasible paths, given the input sequence $X$ and the target sequence $l$, can be denoted as:

\begin{equation}
\begin{split}
    H(\pi|l, X) = & \sum_{\pi \in \mathcal{B}^{-1}(l)} - p(\pi|l, X) \log p(\pi|l, X) \\
    & = - \frac{1}{p(l|X)} \sum_{\pi \in \mathcal{B}^{-1}(l)} p(\pi|X) \log p(\pi|X) \\
    & \quad + \log p(l|X)
\end{split}
\end{equation}

A rapid convergence towards a single feasible path implies a swift reduction in the conditional entropy. By including the term in the objective function that fosters the maximization of the entropy, it subsequently boosts the probability of the paths nearby, enhancing exploration throughout the training process. In this paper, we show that the Maximum Entropy Regularization can benefit the training process for speech recognition.

\subsection{Adaptive Maximum Entropy Regularization}

During the initial phase of CTC training, the randomly initialized network tends to output only blank symbols\cite{peakblank1, peakblank2}. The maximum entropy regularization from the previous section can encourage the network to localize non-blank symbols by penalizing low-entropy blank-only predictions. On the other hand, in the later stage of training, the good and bad paths become distinctive to the model. In this case, the model's prediction should be more deterministic, and thus, fixed entropy regularization can actually act as a hindrance. Therefore, we show the alternative method of entropy regularization, which allows the model to adjust its ambiguity in its prediction.

To begin with, inspired by the improved version of Soft Actor-Critic \cite{SAC2}, we can formulate the following constrained optimization problem:
\begin{equation}
\begin{split}
    \max_{\theta} \log p_{\theta}(l|X) = \max_{\theta} \log \sum_{\pi \in \mathcal{B}^{-1}(l)}p_{\theta}(\pi | X), \\
    \text{ s.t.\ $\mathbb{E}_{p_{\theta}(\pi | X, l)}[- \log p_{\theta}(\pi | X, l)] \geq \mathcal{H}, \forall X$}
\end{split}
\end{equation}
where $\mathcal{H}$ is the desired minimum entropy of the path $\pi$. 

Although our primal CTC objective is not strictly convex, we can approximate the optimization process using a similar Lagrangian dual technique as in \cite{SAC2}. By introducing the Lagrangian dual variable $\beta$, we can formulate the following dual function:

\begin{equation}
    \min_{\beta \geq 0} \max_{\theta} (\log \sum_{\pi \in \mathcal{B}^{-1}(l)}p_{\theta}(\pi | X) + \beta [H(\pi | l, X) - \mathcal{H}])
\end{equation}

The above dual problem can be solved via gradient descent on the learnable parameters $\theta$ and $\beta$. Therefore, we will incorporate this new $\beta$ loss function in addition to the loss function in Eq. \ref{eq.enctc}:
\begin{equation}\label{eq.adamer}
    L_{\beta} = \beta[H(\pi|l, X) - \mathcal{H}]
\end{equation}

The above loss is used to update $\beta$ through the training process. In practical implementation, we use the stop gradient function for both $L_\text{EnCTC}$ and $L_\beta$ since we do not want the gradient of $L_\text{EnCTC}$ to flow through $\beta$ or vice versa. While \cite{SAC2} used dual gradient descent, we jointly trained $\theta$ and $\beta$ for efficient end-to-end training.




\section{Experimental Settings}
\label{sec:experiment_setting}
In this section, we present the benchmarks employed for evaluating our approach, such as the datasets, model, and metrics. 

\subsection{Datasets} LibriSpeech corpus \cite{librispeech} was used for both training and evaluation. LibriSpeech corpus is the widely-used English speech dataset for ASR. For training, we use train-960 subset, which contains 460 hours of clean speech and 500 hours of noisy speech data. For per-epoch evaluation, we use dev-clean set, and for evaluation, we use test-clean and test-other each to compare the performance boost with and without the noise.

\subsection{Model and Metric} In this paper we used Conformer for our baseline model \cite{conformer}. Conformer is the ASR model which combines a convolutional neural network and transformer to capture both local and global dependencies. We implement conformer-small in NVIDIA NeMo toolkit\footnote{{https://developer.nvidia.com/nemo}}, whose encoder comprises 16 conformer layers. The output of the encoder is trained using CTC objective and decoded by greedy searching. For $L_\text{AdaMER}$ and $L_\text{EnCTC}$, we chose initial $\beta$ as 0.2 and 1.0 respectively. Target entropy $\mathcal{H}$ was set to 1.1$U$ where $U$ is the target length. We used AdamW optimizer and Noam learning rate scheduler with 10,000 warmup steps. The models were trained 100 epochs each. Word error rate (WER) is used as the evaluation metric for all experiments. Experiment results are recorded and rendered from WandB dashboard.\footnote{https://wandb.ai/site} 

\section{Results}
\label{sec:results}
\begin{table}[t]
	\centering
	\begin{tabular}{l||c c c }
		\Xhline{2\arrayrulewidth}
        Loss    & test-clean     & test-other  &test-all\\   
		                                   
		\Xhline{2\arrayrulewidth}
	     \begin{small}$L_\text{CTC}$\end{small}     & 5.86 &  14.18  &  10.02  \\
		\begin{small}$L_\text{EnCTC}$\end{small}    &  5.28 &  12.74  &   9.01  \\ 
        \begin{small}$L_\text{AdaMER}$\end{small}    & \textbf{5.13}  &  \textbf{12.62}  & \textbf{8.88}  
		\\\hline
	\end{tabular}
	\caption{WER (\%) (lower is better) on LibriSpeech. test-all denotes the aggregation of test-clean and test-other. The values reported are the average of three runs with different random seeds.}
	\label{tab:libri}
\end{table}

\subsection{Librispeech Results}

Table \ref{tab:libri} demonstrates the Word Error Rate (WER) across different training objectives (i.e., $L_\text{AdaMER}$, $L_\text{EnCTC}$, or $L_\text{CTC}$). We find that using the EnCTC and AdaMER performance improvement is more pronounced compared to standard CTC training for both test-clean and test-other subsets. Specifically, when comparing the WER of CTC with EnCTC and AdaMER, the improvement was 1.01\% and 1.14\%, respectively. This clearly suggests while entropy regularization can help automatic speech recognition tasks, our adaptive entropy regularization can handle it more effectively.

\begin{figure}[t]
	\centering
	\includegraphics[width=1.0\linewidth]{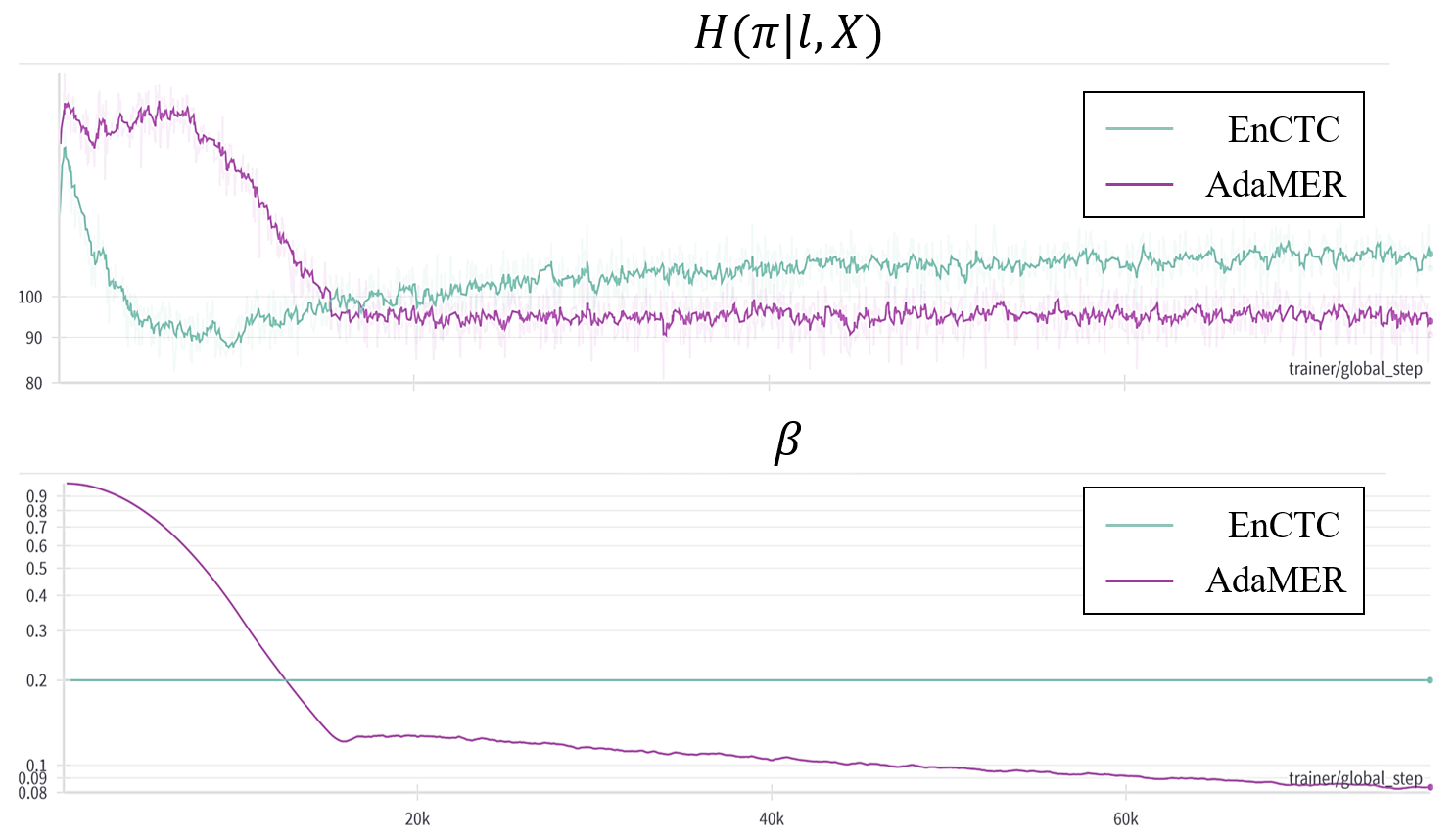}
	\caption{\textbf{Change of $H(\pi|l,X)$ (top) and $\beta$ (down) during training the model with $L_\text{AdaMER}$ (purple) and $L_\text{EnCTC} (green)$}. The horizontal axis is the training step. The results are recorded and rendered from WandB dashboard.} 
	\label{fig:beta}
\end{figure}

Fig. \ref{fig:beta} shows the change of the value of entropy $H(\pi|l,X)$ and the weight $\beta$ throughout training. Notably, the entropy associated with the EnCTC model demonstrates a continual increase over the course of training. In contrast, the AdaMER method consistently maintains the entropy at a relatively stable level. Furthermore, the trajectory of $\beta$ in AdaMER is characterized by a decreasing trend, which persists until the entropy reaches a predefined target value. Beyond this point, the reduction in $\beta$ becomes more gradual. This behavior evidences the adaptive capability of the AdaMER approach in modulating the intensity of entropy regularization.


\section{Conclusions}
\label{sec:conclusions}



In conclusion, our study addresses a pivotal challenge observed in ASR systems trained via Connectionist Temporal Classification (CTC): the issue of overly confident peaky distribution, especially attributed to the blank symbol. Delving into the entropy minimization property inherent to CTC, we recognized the tendency of the model to predominantly predict blank symbols, often leading to sub-optimal outcomes.
We introduced the Adaptive Maximum Entropy Regularization (AdaMER) technique, marking a significant departure from traditional methods that attempted to combine CTC loss with a constant entropy maximization term. AdaMER's uniqueness lies in its dynamic entropy-based scheduler, which modulates the effect of regularization as training progresses. This ensures that while the system is explorative initially, it converges with confident outputs towards the latter stages of training—an essential balance for optimized performance. Our comprehensive experiments demonstrated potential in enhancing the performance of ASR models trained with CTC, especially in tackling the issue of peaky distributions. Our work underscores the importance of adaptive regularization on model confidence in ASR training.

\vfill
\pagebreak
\newpage
\bibliographystyle{IEEEbib}
\bibliography{main}

\end{document}